\documentclass[aps, pra, reprint, longbibliography, superscriptaddress]{revtex4-1}

\usepackage{hyperref}
\usepackage{graphicx}
\usepackage[utf8]{inputenc}
\usepackage{ulem}
\usepackage{xcolor}
\usepackage{amsmath, amssymb}

\begin{document}

\title{Two-dimensional repulsive Fermi polarons with short and long-range interactions}

\author{Raúl Bombín}
\email{raul.bombin@upc.edu}
\affiliation{Departament de F\'{i}sica, Universitat Polit\`{e}cnica de Catalunya, Campus Nord B4-B5, E-08034, Barcelona, Spain}
\affiliation{INO-CNR BEC Center and Dipartimento di Fisica, Universit\`a di Trento, 38123 Povo, Italy}

\author{Tommaso Comparin}
\email{tommaso.comparin@unitn.it}
\affiliation{INO-CNR BEC Center and Dipartimento di Fisica, Universit\`a di Trento, 38123 Povo, Italy}

\author{Gianluca Bertaina}
\affiliation{Dipartimento di Chimica and Dipartimento di Fisica, Universit\`{a} 
degli Studi di Milano, Via Golgi 19, 20133 Milano, Italy}

\author{Ferran Mazzanti}
\affiliation{Departament de F\'{i}sica, Universitat Polit\`{e}cnica de Catalunya, Campus Nord B4-B5, E-08034, Barcelona, Spain}

\author{Stefano Giorgini}
\affiliation{INO-CNR BEC Center and Dipartimento di Fisica, Universit\`a di Trento, 38123 Povo, Italy}

\author{Jordi Boronat}
\affiliation{Departament de F\'{i}sica, Universitat Polit\`{e}cnica de Catalunya, Campus Nord B4-B5, E-08034, Barcelona, Spain}

\begin{abstract}
We study the repulsive polaron problem in a two-component two-dimensional 
system of fermionic atoms. We use two different interaction models: a  
short-range (hard-disk) potential and a dipolar 
potential.  In our approach, all the atoms have the same 
mass and we consider the system to be composed of a uniform bath of a single 
species and a single atomic impurity. We use the diffusion Monte 
Carlo method to evaluate polaron properties  
such as its chemical potential and pair distribution 
functions,  together with a discussion on the deficit of volume induced by the 
impurity. We also evaluate observables that allow us to determine  
the validity of the quasi-particle picture: the quasi-particle residue and the 
effective mass of the polaron. Employing two different potentials allows us to
identify the universality regime, where the properties depend 
only on the gas parameter $n a_s^2$ fixed by the bath density and the two-dimensional scattering length.
\end{abstract}

\maketitle

\section{Introduction}
The polaron problem was put forward by Landau and 
Pekar~\cite{Landau1933,Landau1948} to study the 
interaction of an electron with a crystal lattice. 
In the strongly coupled regime, it was shown 
that the distortion of the lattice, caused by the presence of the electron, may 
induce a local potential that traps the electron. Some years later, Fr\"ohlich 
developed a Hamiltonian formulation~\cite{Frohlich1954} to describe the 
coupling between the impurity (electron) and the phonon modes. Using this 
model, a first variational ground-state solution for the intermediate coupling 
regime was derived by Feynman~\cite{Feynman1955}. Some decades 
later, the picture was completed  with exact results for the Fr\"ohlich model Hamiltonian obtained by 
using the diagrammatic Quantum Monte Carlo (QMC)
method~\cite{Prokofev1998,Mishchenko2000}. The polaron (impurity) problem has 
also  been studied  in other fields of physics such as condensed 
matter ({\it cf.} an impurity of $^3$He in  
bulk $^4$He \cite{Edwards1992,Boronat1999}) 
and nuclear matter~\cite{Bishop1970}.

The achievement of the Bose-Einstein condensate state (BEC) in the past decades 
has provided a new platform for the study of the polaron. The name Bose polaron was coined to indicate an impurity coupled to a BEC, and 
two-component mixtures of ultracold gases featuring a very small concentration of one of the components were proposed as candidate systems where to investigate the 
quasi-particle nature of the impurities~\cite{Tempere2009,Cucchietti2006}. 
In recent years, these configurations have been realized in mixtures of both different hyperfine levels of the same atomic species~\cite{Jorgensen2016} and of
different atoms \cite{Hu2016,Yan2019}. In these experiments, the polaron problem was investigated close to a Feshbach resonance, which allows for the tunability of the interaction strength
between the impurity and the bath. Two branches have been characterized at very low temperatures: the attractive polaron branch, corresponding to the ground state 
of the impurity in the medium, and the repulsive polaron branch, which consists in an excited state of the impurity, where the effective interaction between the impurity and the bath 
is repulsive~\cite{Rath2013,Ardila2015}.

Furthermore, in the context of ultracold gases, Fermi degenerate systems offer 
new possibilities where the polaron picture can arise. Experimental 
measurements have been reported for a spin-down impurity ``dressed" in a bath 
of a spin-up Fermi gas ({\it cf.} in $^6$Li~\cite{Schirotzek2009} ) and for atom mixtures such as $^{40}$K 
impurities into $^6$Li, where attractive and repulsive polaron branches have also been observed~\cite{Kohstall2012}.
While the relation between the bosonic case and 
the Fr\"ohlich formulation is straightforward, the fermionic equivalent 
problem (Fermi polaron) is more challenging and opens the door to a richer scenario. 
Some theoretical works~\cite{Prokofev2008,Nascimbene2009,Massignan2014} have 
studied the polaron  as a first insight into some physical phenomena that are 
characteristic of the strongly interacting regime: the pairing mechanism that 
gives rise to the BEC-BCS crossover 
\cite{Zwierlein2006,Partridge2006,Schirotzek2009}, possible itinerant 
ferromagnetism in two-component systems~\cite{Pilati2010, 
Jo2009,Valtolina2017,Comparin2018} or the Kondo effect in systems containing magnetic 
impurities~\cite{Nakagawa2015}.

The realization of quantum  degenerate systems composed of atoms with large 
magnetic moment has motivated additional interest in the polaron problem. The 
dominant dipolar interaction between these atoms is of longer range and 
anisotropic. This was first achieved with
  Cr atoms~\cite{Stuhler2005,Griesmaier2005} and more recently
also with Dy~\cite{Lu2011,Lu2012} and 
Er~\cite{Aikawa2012,Aikawa2014} that have a larger magnetic moment than Cr. 
Regarding the polaron problem, the report of experimentally accessible ultracold 
mixtures of Er and Dy~\cite{Trautmann2018} and the study of low 
concentration impurities of $^{163}$Dy in a $^{164}$Dy 
droplet~\cite{Wenzel2018} have motivated the study of the dipolar polaron in 
three~\cite{Kain2014} and in quasi-two dimensional configurations 
\cite{Ardila2018}. The dipolar polaron has also been studied in a bilayer 
geometry, where localization effects are predicted near the 
crystallization point~\cite{Matveeva2013}. 

In two dimensions (2D), quantum correlations are enhanced compared to the three-dimensional 
(3D) case. While the one particle-one hole picture has demonstrated 
its utility to study the Fermi polaron problem in 3D systems \cite{Prokofev2008}, it fails when trying to 
accurately reproduce the physics of the equivalent system in 2D~\cite{Vlietinck2014}. Up 
to now, some efforts have been put in the study of the repulsive Fermi polaron, 
studied as the repulsive branch of a system with short-range interactions 
({\it cf.}  Refs.~\cite{Koschorreck2012,Frohlich2011}  and 
\cite{Schmidt2012,Ngampruetikorn2012,Vlietinck2014} for experiment 
and theory, respectively). However, the equivalent system 
but with dipolar interactions, which in principle would be accessible 
in current experiments, remains unexplored. 

 In this work, we study the repulsive Fermi polaron of a two-component system, 
labeled as $\uparrow$ and $\downarrow$ in analogy with 
spin-1/2 particles. The system, consisting of $N =  N_\uparrow +1$ particles, 
contains a single atomic impurity immersed in a bath composed of 
 $N_\uparrow$ atoms. We study this model with two different types of 
interparticle interaction, which allows us to determine the universality of the 
system. The first model considers that the only interaction present in the 
system is a short-range one between the up and down particles, modeled as a 
hard-disk potential, while the bath is considered to be an ideal Fermi gas. The 
second one assumes dipolar interactions between all the particles. In the latter 
case, we assume that all the dipolar moments are polarized along the 
direction perpendicular to the plane of motion, so that the 
interaction between them is isotropic (see for instance
Ref.~\cite{Comparin2018}).

The paper is organized as follows. In Sec.~\ref{sect:model}, we report 
the two models that we use to describe the two-dimensional system composed of 
a single impurity in a polarized Fermi bath. Sec.~\ref{sect:method} 
discusses  the  diffusion Monte Carlo method 
that we use for the calculations. In Sec.~\ref{sec:results}, we show our 
QMC results for the polaron energy and the pair correlation function 
between the bath and the impurity, together with an analysis of the deficit of 
volume induced by the impurity. We also validate the 
quasi-particle picture by evaluating the quasi-particle residue and the 
effective mass of the polaron. Finally, Sec.~\ref{sect:conclusions} contains 
the main conclusions of our work, emphasizing the limits of universality of the 
2D Fermi polaron problem.  


\section{Models}
\label{sect:model}
We describe a two-component Fermi system in 2D.
The system is composed by  $N = N_\uparrow + 1$ atoms of equal mass $m$,
representing a single-component bath with one additional atomic impurity.
To reproduce the physics of a uniform infinite system, we put all the  
particles in a square box with periodic boundary conditions,
with the box side $L$ fixed by the density $n$ of the bath ($L=\sqrt{N_{\uparrow}/n}$).  
The $N$-particle Hamiltonian reads
\begin{equation}
\hat{H} =
-\frac{\hbar^2}{2m}  \nabla_\downarrow^2
-\frac{\hbar^2}{2m} \sum_{i=1}^{N_{\uparrow}} \nabla_i^2
+ \sum_{i<j}^{N_{\uparrow}} V^{\mathrm{bath}}(r_{ij}) +
\sum_{j=1}^{N_{\uparrow}}V^{\mathrm{int}}(r_{\downarrow j}),
\label{eq:Hamiltonian}
\end{equation}
where $r_{ij} \equiv |\mathbf{r}_i - \mathbf{r}_j|$ is the distance 
between two bath particles and $r_{\downarrow j} \equiv |\mathbf{r}_\downarrow - \mathbf{r}_j|$ is
the distance between a bath particle at $\mathbf{r}_j$ and the impurity position
$\mathbf{r}_\downarrow$.
Throughout this work, labels $i$ and $j$ refer to 
bath particles.
$V^{\mathrm{bath}}(r)$ is the two-body potential between the 
bath particles, and $V^{\mathrm{int}}(r)$ is the interaction potential between 
the impurity and the bath. In the following, we describe the two different 
interaction models that we study.

\subsection{Hard-disk Model}
\label{subsect.HD}

We first consider a hard-disk model for the repulsive polaron, which is 
experimentally relevant for the description 
of the upper metastable branch of the Fermi polaron~\cite{Pilati2010}. 
In this case, the bath is non-interacting 
($V^\mathrm{bath}(r)= 0$) and the impurity interacts with the bath particles with 
a hard-core potential,
\begin{equation}
V^{\mathrm{int}}(r)= \left\{ \begin{array}{lcc}
             \infty & r \leq R \\
             \\0 & r > R .\\
             \end{array}
   \right.
   \label{eqn.potHD}
\end{equation}

It is important to recall that, in 2D, the scattering amplitude 
depends logarithmically on momentum, so that the definition of the scattering length 
$a_s$ involves an arbitrary constant. Two alternative conventions are typically 
used. In the first one, $a_s$ is defined to fulfill $a_s=R$ for a hard-core 
potential, so that the two-body scattering wave function vanishes at $r=a_s$ 
\cite{Pilati_QuantumMonteCarlo_2005} in analogy with the 3D case. This is 
the convention 
that we use in this work. With such definition, the two-body binding energy for
an attractive contact interaction is 
$|\epsilon_b|=4\hbar^2/(ma_s^2e^{2\gamma})$, with $\gamma\simeq 0.577$ Euler's
constant~\cite{Bertaina_BCSBECCrossoverTwoDimensional_2011,Bertaina2013}. Another
definition of the 2D scattering length (now indicated by $b$) aims at 
maintaining a simple relation with the binding energy 
$|\epsilon_b|=\hbar^2/(mb^2)$, in analogy with the 3D attractive 
problem \cite{Petrov2001,Schmidt2012}. The relation between the two conventions 
is $b=a_se^{\gamma}/2$.

For the hard-disk model, all the physics in the system is condensed into the 
gas parameter $na_s^2$. We also notice that the closer $na_s^2$ is to 
unity, the less this model is expected to faithfully describe the repulsive branch 
of the polaron, since coupling to molecular states is 
completely ignored.

\subsection{Dipolar Model}
In the second model, all the particles in the system 
interact with each other through the same dipolar potential. We also consider all the 
dipoles to be polarized in the direction perpendicular to the plane of motion, so that 
the interaction between them is isotropic.  Thus, the two potentials appearing 
in the Hamiltonian of Eq. (\ref{eq:Hamiltonian}), take the form:
\begin{equation}
V^{\mathrm{bath}}(r) = V^{\mathrm{int}}(r) = \frac{C_{dd}}{4\pi}\frac{1}{r^3}.
\label{eq:potdip}
\end{equation}

The purpose of including dipolar interparticle
interaction between all the particles in the system, and not only between the bath and the 
impurity, is to study an experiment that could be suitable for current
state-of-the-art experiments:  A polarized system of fermionic polar atoms (such as $^{161}$Dy or $^{167}$Er \cite{Trautmann2018}), tightly confined in the polarization direction, with the majority of spin-up atoms forming the bath, and a vanishingly small concentration of spin-down impurity atoms. It would also be a good model in the case in which the impurity is an isotope of the same element, such as: $^{162}$Dy into a bath of $^{161}$Dy. Differently from the short-range case, in the dipolar case the interaction between uneven fermions cannot be neglected.

For a dipolar system, the Hamiltonian  (\ref{eq:Hamiltonian}) can be 
written in dimensionless form by expressing all distances in units of the 
characteristic length $r_0=m C_{dd}/(4\pi \hbar^2)$ and energies in units of 
$\epsilon_0=\hbar^2/m r_0^2$. Hence, 
properties of the homogeneous system are governed by  the dimensionless density 
$nr_0^2$ encoding the strength of the interactions, as it was done in previous 
works~\cite{Astrakharchik2007,Macia2011,Comparin2018}. Although in three dimensions 
the dipole-dipole potential is long ranged, in two dimensions it is not. 
Therefore, in the low-density regime it can be reduced to a contact interaction 
and we can use the gas parameter $na_s^2$ for a better 
comparison with other potentials, such as the one in the previous subsection.
In dipolar units, the scattering length has the value $a_s 
= r_0 e^{2\gamma}$
\cite{Macia2011,Ticknor2009}, so that $n a_s^2 \simeq 10\,n r_0^2$. 

\section{Method}
\label{sect:method}

We employ the diffusion Monte Carlo (DMC) method~\cite{Hammond1994,Kosztin1996} 
for finding 
the ground state of the Hamiltonian in Eq.~\eqref{eq:Hamiltonian}. The 
DMC algorithm is a stochastic method that allows us to find the ground state of 
the system by propagation in imaginary time. For bosonic systems it gives exact 
results, within statistical errors,  when the imaginary-time step tends to 
zero,  $\delta\tau \rightarrow 0 $,  and for an infinite population of walkers 
$N_{W} \rightarrow \infty$ (a walker being a set of $N$ coordinates). In 
practice, convergence can be achieved using small imaginary-time steps and 
large enough number of walkers. It is well known that a trial wave function 
$\Psi_T$ used for importance sampling reduces the variance without introducing 
any additional bias, as long as $\Psi_T$ has a finite overlap with the exact 
ground-state wave function of the system. For fermionic systems, the wave 
function is not positive definite, giving rise to the so called sign problem. The 
most common technique to keep this problem under control is 
the fixed-node approximation.  In this scheme, 
the DMC method is exact only if the  nodal surface of $\Psi_T$ coincides with 
the one of the ground state. Otherwise, it becomes a variational method,
which provides an upper bound for the ground-state energy.

In our simulations, we use a Jastrow-Slater trial wave function,
\begin{equation}
\Psi_T(\mathbf{R}) = \Psi_A(\mathbf{R}_\uparrow)
\, \Psi_J(\mathbf{R}) \ ,
\label{jastrow-slater}
\end{equation}
where 
$\mathbf{R} = \lbrace \mathbf{r}_1, \dots,
\mathbf{r}_{N_\uparrow}, \mathbf{r}_\downarrow \rbrace$ is the set of all
$N_\uparrow + 1$ particle coordinates, and
$\mathbf{R}_\uparrow$ is restricted to the $N_\uparrow$ particles of the bath.
The antisymmetric wave function for the bath,
$\Psi_A(\mathbf{R}_\uparrow)$,
is a Slater determinant, where we use plane waves
as single-particle orbitals.
These orbitals, which correspond to  the nodal surface of the free 
Fermi gas, are accurate enough for the low 
densities considered here, as it was shown in a recent work~\cite{Comparin2018}.

The symmetric Jastrow part is written as
\begin{equation}
\Psi_J(\mathbf{R}) = \prod_{j=1}^{N_\uparrow} f_{\uparrow\downarrow}(r_{j\downarrow}) \, 
\prod_{i<j}^{N_\uparrow} f_{\uparrow\uparrow}(r_{ij}) \ .
\label{jastrow}
\end{equation}
The two-body correlation functions $f_{\uparrow\uparrow}(r)$  and $f_{\uparrow\downarrow}(r)$ are constructed from the zero-energy two-body solution satisfying 
the conditions  $f_{\uparrow\uparrow}(L/2)=f_{\uparrow\downarrow}(L/2) = 1$, $f'_{\uparrow\uparrow}(L/2)=f'_{\uparrow\downarrow}(L/2) = 0$. For the dipolar model, the 
two-body solution is matched at a certain distance $r_{M}$ with a 
symmetrized phononic tail, reproducing the long distance behavior in the 
medium~\cite{Astrakharchik2007}. In general, the bath/bath and bath/impurity
correlations
 are significantly different, so that
in the dipolar 
model we consider different values of
the matching distance $r_{M}$ for the two cases. Therefore we have
two variational parameters:
$r_{M}^{\uparrow\uparrow}$ and $r_{M}^{\uparrow\downarrow}$. For the 
hard-disk model, Jastrow correlations  are  implemented only for the 
impurity-bath pairs, since the bath is non-interacting ($f_{\uparrow\uparrow}(r)=1$). In the 
latter case, the only variational parameter is $r_{HD}\le L/2$, at which we impose the 
conditions $f_{\uparrow\downarrow}(r_{HD}) = 1$, $f_{\uparrow\downarrow}^\prime(r_{HD}) = 0$.

In a DMC calculation, expectation values of a given operator $\hat{O}$ are obtained by 
sampling over the mixed probability distribution $f(\mathbf{R},\tau) = 
\Psi_T(\mathbf{R})\phi(\mathbf{R},\tau)$. For a system of bosons, $\phi(\mathbf{R},\tau)$ is the exact wave function of the system, while, for a fermionic system, it corresponds to the Fixed-Node upper bound related to the choice of the nodal surface. For long enough 
imaginary time, components of $\phi$ that are orthogonal to the ground state 
$\phi_0$ are removed and the only relevant contribution comes from $\phi_0$,
\begin{equation}
\langle\hat{O}\rangle_\mathrm{DMC}=\frac{\langle\Psi_T\vert\hat{O}\vert\phi_0\rangle}{\langle\Psi_T\vert\phi_0\rangle} =
\lim_{\tau\rightarrow\infty}\frac{\int d\mathbf{R} \;\phi(\mathbf{R},\tau) \hat{O}
\Psi_T(\mathbf{R})}{\int d\mathbf{R}\; \phi(\mathbf{R},\tau) \Psi_T(\mathbf{R})}. 
\label{eq.DMCest}
\end{equation}
Equation (\ref{eq.DMCest}) gives unbiased results when the operator $\hat{O}$ 
is the Hamiltonian or it commutes with $\hat{H}$.  
For diagonal operators that do not commute with the Hamiltonian, it is still 
possible to obtain exact values using the pure estimators technique 
\cite{Casulleras1995}. In the case of non diagonal operators, obtaining a pure 
estimator is more subtle. In this work, we will restrict our results to a first 
order correction in $\Psi_T$ given by the extrapolated estimator,
\begin{equation}
\langle\hat{O}\rangle \simeq 
2\langle\hat{O}\rangle_{\mathrm{DMC}} - \langle\hat{O}\rangle_{\mathrm{VMC}},
\label{eq.DMC_extrap}
\end{equation}
{where $\langle\hat{O}\rangle_{\mathrm{VMC}}=\frac{\langle\Psi_T\vert\hat{O}\vert\Psi_T\rangle}{\langle\Psi_T\vert\Psi_T\rangle}$
is the variational Monte Carlo (VMC) estimator. The above extrapolation is accurate when the DMC correction to 
the VMC result is small.

\section{Results}
\label{sec:results}

The QMC results that appear in this section will be compared 
with two 
approximate theories to benchmark them. This will be also of some utility to 
study the regime in which the system becomes universal in terms of the gas 
parameter. As a first approximation, we will compare our energies with the 
prediction that mean-field theory offers for the system~\cite{Schick1971}. On 
the other hand, we will also compare our results with a T-matrix study of 
the repulsive Fermi polaron~\cite{Schmidt2012}. The authors of Ref.~\cite{Schmidt2012} considered the ultradilute limit of spin-up 
impurities immersed in an spin-down bath, which is treated as an ideal Fermi 
gas. Quasi-particle properties (effective mass and quasi-particle residue) were 
then evaluated both for the attractive and the repulsive branches of a 
system where the impurity interacts with the bath via a short-range potential
having scattering length $a_s$. Due to the similarity of the repulsive branch 
studied in that model with our  hard-disk system described in 
Sec.~\ref{subsect.HD}, it is worthy to compare it with the results obtained 
with our QMC approach.
 
\subsection{Energy of the polaron}
\label{subsec.energy}
The energy of the polaron is an important and experimentally accessible 
observable. 
It is defined as the energy difference between the pure system of $N_\uparrow$
particles
and the same system with an added impurity, at fixed volume. Making use of this 
definition it can be directly evaluated in QMC simulations as the chemical 
potential of the impurity, that is, extracted from the difference between the energy
of the system with an added impurity $E (N_\uparrow, 1)$, and the one of the pure system $E(N_\uparrow,0)$ at fixed volume:
\begin{equation}
\varepsilon_p = \left[E (N_\uparrow, 1) - E(N_\uparrow,0)\right]_V \ .
\label{eq:ener_pol}
\end{equation}
In mean-field theory an expression for the 2D polaron energy can be obtained, valid in the limit of vanishing density
\begin{equation}
\varepsilon_{\mathrm{MF}} = \frac{4 \pi\hbar^2 n }{m\ln(c_0na_s^2)} 
\ .
\label{eq:ener_polMF}
\end{equation}
The dependence of the mean-field prediction (\ref{eq:ener_polMF})  on 
a free  parameter $c_0$ is a peculiarity of 2D  systems that is related to the 
features of scattering theory in 2D~\cite{Pitaevskii2016}.
This free parameter is related to 
a characteristic energy scale of the system~\cite{Bertaina2013,Comparin2018}. 
In the present work, we set it to the value $c_0=e^{2\gamma} \pi/2 \simeq 
4.98$, corresponding to using an energy scale equal to the Fermi energy
$E_\mathrm{F} = 2 \hbar^2 \pi n / m$.

In Fig.~\ref{fig:polaron_ener}, we show our QMC results compared to 
the 
mean-field prediction  of Eq.~(\ref{eq:ener_polMF}).  We plot 
the polaron energy in units of the mean-field energy, so that deviations 
from mean field are enhanced.  Although being a good approximation, mean 
field fails to accurately reproduce even the lower densities considered in this 
work, which is a well known fact in two-dimensional 
gases~\cite{Astrakharchik2009}. As the density is increased, the mean-field 
prediction has a logarithmic divergence and thus it does not stand as a good 
energy scale for values of $na_s^2 >10^{-3}$. For this reason, in 
the inset of Fig.~\ref{fig:polaron_ener} we plot the polaron energy,  for 
the highest gas parameters, in units of the Fermi 
energy, $E_F$. The error bars that appear in 
Fig.~\ref{fig:polaron_ener} include both statistical and systematic errors, 
 the latter being the largest contribution.  In the low density regime,  the 
systematic error is dominated by the finite value of the imaginary-time step
$\delta\tau$, while
for the higher densities the main source of error comes from 
finite-size effects. Concerning this latter issue, calculations have been 
done using 61 bath particles for all the dipolar system, while, for the hard-disk model, the exclusion of volume caused by the impurity makes it necessary to include 121 particles in the bath to maintain finite size effects under control when the gas parameter is higher than $na_s^2 \geq 10^{-2}$. In the case of hard-disk interaction, systematic errors for the polaron energy are of
the order of  0.5\%, while for dipolar systems they grow up to 1\%. 
 
\begin{center}
\begin{figure}[h]
\includegraphics[width=1.02\linewidth]{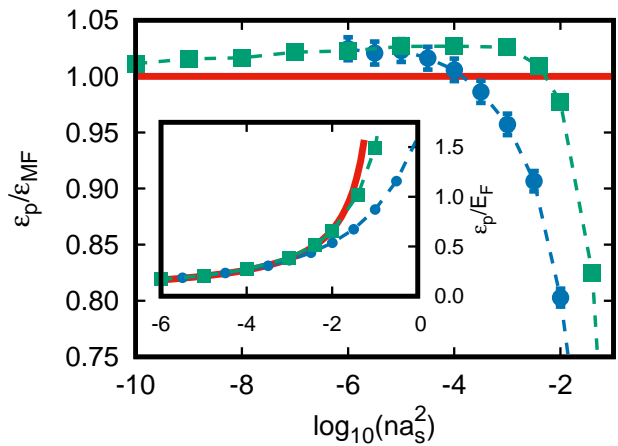}
\caption{Energy of the polaron in units of the mean-field energy in
Eq.~\eqref{eq:ener_polMF}. 
The red line is the mean-field prediction, while green and blue symbols are DMC 
results for hard-disk and dipolar models, respectively. Dashed lines are guides to the 
eye. For large values of the gas parameter, the mean-field energy is not a good energy 
scale due to the logarithmic divergence of Eq. \eqref{eq:ener_polMF}. Inset: polaron energy, in units of the bath Fermi energy $E_\mathrm{F}$, plotted for larger values of na$_s^2$.}
\label{fig:polaron_ener}
\end{figure}
\end{center}

\subsection{Pair distribution function}
The presence of the impurity affects the local properties of the bath.
This effect can be analyzed by looking at the pair 
distribution function between the background and the impurity 
$g^{\uparrow\downarrow}(r)$, sometimes referred to as the
density profile of the bath around the impurity. In DMC simulations, we can 
evaluate both this distribution function and the one involving bath 
particles,  $g^{\uparrow\uparrow}(r)$,
\begin{align}
g^{\uparrow\uparrow}(r) &= \frac{2}{n N_\uparrow}\frac{\int 
d\mathbf{R} \, \phi_0(\mathbf{R}) \Psi_T(\mathbf{R}) \,    
\sum_{i<j}^{N_\uparrow}\delta (r- r_{ij})}
{\int d\mathbf{R} \,
\phi_0(\mathbf{R}) \Psi_T(\mathbf{R})},
\label{eq.gr}
\\
g^{\uparrow\downarrow}(r) &= \frac{1}{n}\frac{\int 
d\mathbf{R} \, \phi_0(\mathbf{R})  \Psi_T(\mathbf{R}) \,  
\sum_{j=1}^{N_\uparrow}\delta (r-r_{\downarrow j})  
}{\int d \mathbf{R} \, \phi_0(\mathbf{R}) \Psi_T(\mathbf{R})}.
\label{eq.denprof}
\end{align}
Figure~\ref{fig:denprof} shows  $g^{\uparrow\downarrow}(r)$, as a function 
of the dimensionless quantity $r\sqrt{n}$, for different gas parameters 
and for the two models considered in this work. The plot indicates that the hole 
around the impurity, arising from repulsive correlations between the impurity and bath particles, 
grows when the gas parameter is increased. We also notice that, at the lowest interaction strength 
shown for the dipolar model ($na_s^2\simeq10^{-4}$), the distribution function closely resembles the one of the hard-disk model (except at distances compared to the core radius $R=a_s$) indicating the approaching to the low-density universal regime, similar to what one finds when comparing the polaron energies for the two models. For the dipolar model, the radial distributions have been evaluated using the pure estimators technique~\cite{Casulleras1995} whilst for the hard-disk model, DMC results have been extrapolated as it was explained in the previous section (see Eq.~\eqref{eq.DMC_extrap}). This also applies for the data in Fig. \ref{fig:def_vol}, and in both cases error bars are chosen to cover systematic errors.

\begin{figure}[h]
\begin{center}
\includegraphics[width=8.5cm]{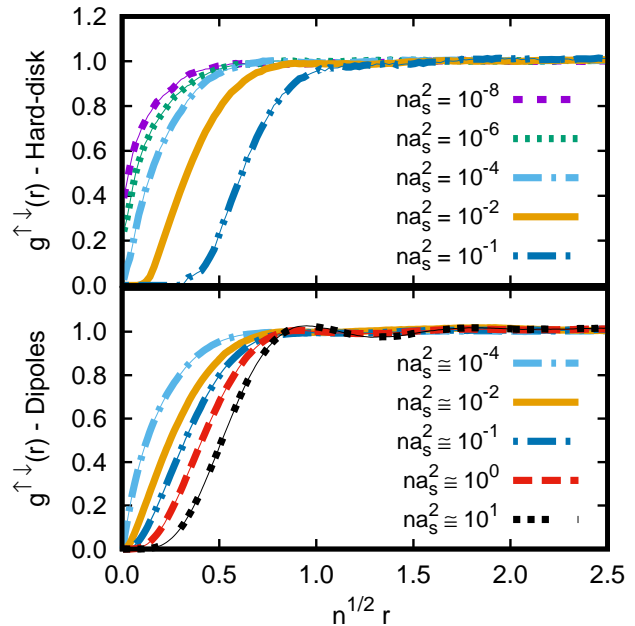}
\caption{Monte Carlo results for the  pair distribution function
$g^{\uparrow\downarrow}(r)$ between the 
impurity 
and the bath, evaluated for different values of the 
gas parameter $na_s^2$ for the hard-disk model (top panel) and for the dipolar 
one 
(bottom panel).}
\label{fig:denprof}
\end{center}
\end{figure}

As the gas parameter approaches $n a_s^2 \simeq 1$, the radius of the
hard-disk model starts to approach the mean interparticle distance and
the model ceases to capture the physics of the repulsive branch with
short-range interactions.  {Instead, the dipolar model still maintains
  its physical meaning in the high-density regime and
  $g^{\uparrow\downarrow}(r)$ features Friedel oscillations,
  indicating the formation of shells of particles around the impurity.
  It is worth mentioning that all the radial distributions shown in
  Fig.~\ref{fig:denprof} are evaluated in a system containing 61 bath
  particles except for the two highest densities shown for the
  hard-disk model ($na_s^2=10^{-2}$ and $10^{-1}$). In these latter
  cases, the large amount of volume excluded by the impurity enhances
  the finite-size effects and the use of 121 bath particles is needed
  to keep them under control.

Due to the interaction between the impurity and the medium as well as
the statistics of the particles in the bath, the volume occupied by
the impurity is different from the one of any of the bath particles.
If one considers a mixture with a very low concentration of
impurities, the total density of the mixture can be written at fixed
pressure $P$ as $\rho(P,x) = \rho(P,x=0) (1+\alpha x)^{-1}$ with $x$
the concentration of the impurity atoms and $\alpha$ the excess volume
parameter. As it was shown in~~\cite{Saarela1993,Kurten1983},
in the limit $x \to 0$, $\alpha$ can be approximately evaluated from
the $k=0$ value of the static structure factor
$S^{\uparrow\downarrow}(k)$ correlating the impurity and the bath
particles:
\begin{equation}
S^{\uparrow\downarrow}(0) = -(1+\alpha) \ ,
\label{eqn:def_vol}
\end{equation}
where $S^{\uparrow\downarrow}(k)$ is related to the Fourier transform of the 
radial distribution function $g^{\uparrow\downarrow}(r)$ 
\begin{equation}
S^{\uparrow\downarrow}(k) =
n \int d\mathbf{r}\,
e^{i\mathbf{k} \cdot \mathbf{r}}
\left(g^{\uparrow\downarrow}(r)-1 \right).
\label{eqn:structre_factor}
\end{equation} 

The sign of $\alpha$ carries information on whether there is an excess or  
deficit of
volume induced by the inclusion of the impurity particle in the bath: $\alpha > 
0$  ($\alpha<0$)  indicates that the impurity occupies more (less) 
volume than a given bath particle. This quantity has been evaluated in condensed-matter 
systems, for example for an $^3$He atom in bulk $^4$He. There, it was 
shown that the $^3$He atom occupies near 30\% more volume than the average 
volume occupied by the particles of the $^4$He bath~\cite{Boronat1999}. In 
that 
case, the increase of volume can be qualitatively explained in terms of the 
different zero-point motion that the two isotopes have, stemming from the 
mass difference. 

For a 
system where all atoms have the same mass and the same interparticle interaction 
but where the species are distinguished by their spin component, as it is the 
case of our dipolar system, a decrease of volume would arise because of Fermi 
statistics. In order to quantify this reduction, we 
evaluate the impurity-bath static 
structure factor of Eq.~\eqref{eqn:structre_factor} for our system of dipoles at 
different densities (see bottom panel of Fig.~\ref{fig:def_vol}). For this model, the volume coefficient $\alpha$ is negative for all the range of densities that we analyze, telling us 
that the impurity occupies less volume than one of the bath particles, since
these are pushed further apart from each other due to Fermi repulsion.
We see that
the excess volume $\alpha$ decreases in magnitude with increasing 
density, that is, when the potential contributions to the energy start to be 
important compared to the Fermi repulsion. If one keeps increasing the density 
of the system up to the crystallization point ($nr_0^2\sim 50$ \cite{Matveeva2012}), the volume 
coefficient would approach zero ($\alpha \rightarrow 0 $), as it would be for 
an impurity which is barely distinguishable from the bath atoms.

The deficit of volume can also be analyzed in our hard-disk model (see top panel of Fig.~\ref{fig:def_vol}). In this case, however, the physics is different from the dipolar model, where the only difference between the two species comes from Fermi statistics. In this model one has also to consider that the only interaction present in the system is that of the impurity with the ideal Fermi bath. As a result, two effects compete and dominate over each other in different regimes. For low values of the gas parameter, where the hard-core radius is small compared to the mean interparticle distance, one expects that all the deficit of volume would be caused by Fermi statistics, similar to the dipolar case. This is what can be seen when comparing the QMC results for the two models in Fig.~\ref{fig:def_vol}: up to values of $na_s^2\le$10$^{-4}$, the two interactions potentials give the same $\alpha$ parameter. On the contrary, as the gas parameter increases and the system abandons the universal regime, the radius of the hard-core starts to be compatible with the interparticle distance and $\alpha$ is greater than that from the equivalent dipolar system. It is worth noticing that for the highest gas parameter considered for this model, $na_s^2$=10$^{-1}$, the volume coefficient becomes positive, meaning that the impurity, in this regime, occupies a bigger volume than an average particle in the ideal Fermi bath considered.  
\begin{figure}[h]
\includegraphics[width=8.5cm]{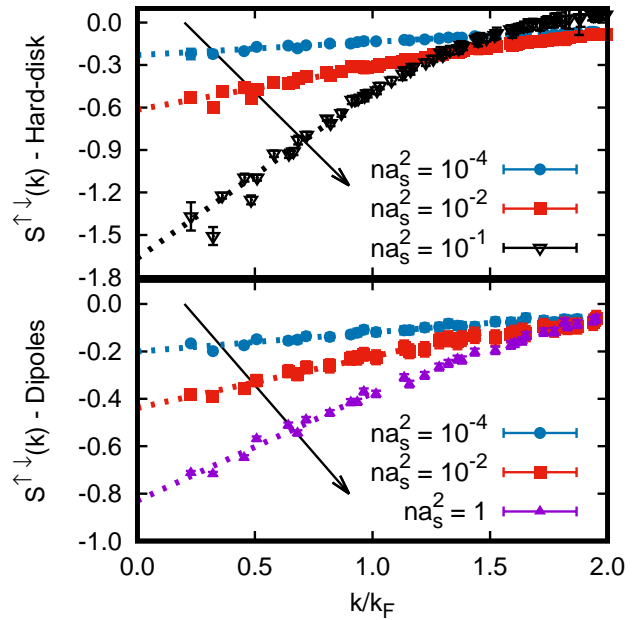}
\caption{Static structure factor $S^{\uparrow\downarrow}(k)$ involving correlations between the impurity and the bath particles, for small values of $k/k_F$ with $k_F = \sqrt{4\pi n}$ the Fermi momentum. Top (bottom) panel: results 
correspond to the hard-disk (dipolar) system for different values of the gas parameter. Same color and symbols are used to emphasize when the two models are evaluated at the same gas parameter. Dashed lines correspond to a linear extrapolation to $k\to 0$. The arrows indicate
increasing density.}
\label{fig:def_vol}
\end{figure}

\subsection{Quasi-particle properties}
\label{subsec.Z}

In the weakly-interacting regime, one can assume that the wave function $\phi$ describing 
the   state of the bath plus the impurity system has an important overlap with the
state $\Phi ^{\mathrm{NI}}$ in which interactions between the impurity and the bath are absent.
The latter is a state representing a system containing a non-interacting impurity with 
momentum $k = 0$ immersed in an unperturbed single-component bath.
The quasi-particle residue $Z$ is defined through this overlap \cite{Vlietinck2013}:
\begin{equation}
Z = \left|\langle \Phi ^{\mathrm{NI}}|\phi\rangle\right|^2 \;.
\label{eqn:Z_def}
\end{equation}
For the system with hard-disk interaction, where the bath is an ideal 
Fermi gas, 
$\Phi ^{\mathrm{NI}}$ reduces to $|\mathrm{FS} + 1\rangle $, which
stands for a Fermi sea with an added non-interacting impurity at zero
momentum. In our dipolar model, in contrast, bath particles interact with each 
other, so that $\Phi ^{\mathrm{NI}}$ is the state of the interacting bath with the 
addition of a non-interacting impurity at zero momentum.
The quasi-particle residue in Eq.~\eqref{eqn:Z_def} also represents the probability of free propagation of 
the impurity in the medium.  

In the theory of Fermi liquids, the quasi-particle residue $Z$ corresponds to 
the 
jump 
in the momentum distribution $n(k)$ at the Fermi momentum. In our study, if we consider 
the impurity as the zero-density limit of a Fermi sea, we obtain the relation $Z 
= n_{\downarrow}(k=0) - n_{\downarrow}(k=0^+)$. The components at $k>0$ scale with the inverse 
volume, so that they are negligible in the thermodynamic limit for the bath~\cite{Punk2009,Guidini2015}. In QMC simulations 
in real space, the quasi-particle residue is best extracted from the Fourier 
transform of the momentum distribution, the one-body density matrix (OBDM). 
While its integral over volume would yield $n_{\downarrow}(k=0)$ for a finite system, 
its asymptotic value at $r\to L/2$ is a better estimate of $Z$,  since the finite-size component is automatically removed.
Following this scheme, we evaluate the quasi-particle residue  from the asymptotic 
behavior of 
the OBDM involving the impurity, which, in the DMC framework is evaluated 
from the following estimator
\begin{equation}
Z=\lim_{|\mathbf{r}^\prime_\downarrow-\mathbf{r}_\downarrow|\rightarrow L/2}
\left\langle \frac{\Psi_T(\mathbf{R}_{\uparrow},\mathbf{r}_{\downarrow}
^ { \prime } ) } { \Psi_T(\mathbf{R}_ { 
\uparrow},\mathbf{r}_{\downarrow})}
\right\rangle.
\label{eq.Z_residue}
\end{equation}

Since this DMC estimator is non-diagonal, the result is
generally biased due to the choice of the trial wave function ({\it cf.}
section \ref{sect:method}). 
Our estimation is based on the extrapolated estimator in Eq.~(\ref{eq.DMC_extrap}) 
which we expect to be accurate enough due to the quality of the trial wave 
function, especially at low densities.
In Fig.~\ref{fig:Z_residue}, we show our results  for the residue $Z$,  
following the 
prescription of Eq.~(\ref{eq.Z_residue}), both for hard disks and dipoles.  We 
find that a universal regime can be identified for gas parameters lower than $na_s^2<10^{-3}$, up to where relative differences between the quasiparticle residues evaluated for the two models remain below 5\%.  This relative deviations are comparable to the ones reported for the polaron energy at that same gas parameter in Sec.~\ref{subsec.energy}.  However, in the regime $na_s^2>10^{-3}$, clear differences  between the two models appear: for the 
dipolar model the quasi-particle residue features 
values higher than $0.6$ in 
all the interval of $na_s^2$ considered here. On the contrary, for the  
hard-disk model, $Z$ is highly suppressed as 
the gas parameter is increased, reflecting that the interaction radius begins to be comparable to the 
interparticle distance, making it difficult for the impurity to perform a free 
displacement. Noticeably, for the largest value of the gas parameter ($na_s^2=4\cdot 10^{-1}$)
the residue almost vanishes suggesting a tendency of the impurity to get localized as the interaction strength becomes very large.
In the same plot, we include the T-matrix results from 
Ref.~\cite{Schmidt2012}, corresponding to the quasi-particle residue of 
the repulsive 
branch of the 2D Fermi impurity problem with short-range interactions. 
These results are in 
reasonable agreement with our hard-disk impurity model, up to a regime where the excited repulsive polaron loses its identity due to strong 
coupling to the molecular branch.
\begin{figure}[h]
\includegraphics[scale= 1]{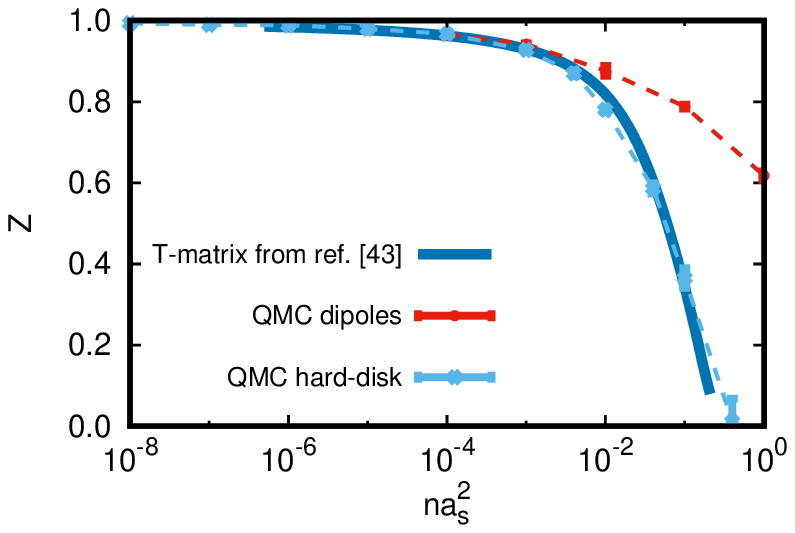}
\includegraphics[scale= 1]{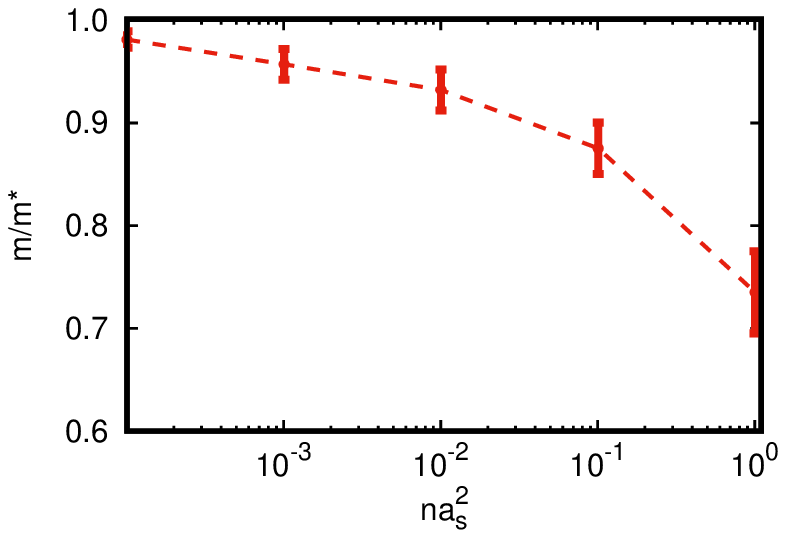}
\caption{Quasi-particle residue $Z$ (top panel) and effective mass of the 
polaron 
(bottom panel) as a 
function of the gas parameter $na_s^2$. Red symbols correspond to he dipolar 
system, blue ones correspond to the hard-core impurity, and the solid blue line 
shows the many-body T-matrix theory results~\cite{Schmidt2012}. Dashed lines are guides to the eye.}
\label{fig:Z_residue}
\end{figure}

The other quantity that is of relevance for studying the polaron in a 
quasi-particle 
picture is its effective mass, that is the mass of the 
quasi-particle formed by the impurity ``dressed" by the medium. In a DMC simulation, the 
effective mass $m^*$ is obtained from the asymptotic diffusion coefficient in 
imaginary time of the 
impurity throughout the bath~\cite{Boronat1999,Ardila2015},
\begin{equation}
 {m \over m^*} = \lim_{\tau\to\infty} {1\over 4\tau}{D_s^{\downarrow}(\tau) \over D_0},
\end{equation}
with $D_0 = {\hbar^2 \over 2m}$ being the free-particle diffusion constant and 
$D_s^{\downarrow}(\tau) = \langle
(\mathbf{r}_\downarrow(\tau)- \mathbf{r}_\downarrow(0))^2\rangle$ the squared 
imaginary-time displacement of the impurity.  In the bottom panel of
Fig.~\ref{fig:Z_residue} we report our DMC results for the dipolar system, which show that interaction effects increase the effective mass of the polaron by roughly 30\% as the gas parameter 
increases up to $na_s^2\sim1$. 
Although the effective mass for the hard-disk model has not been evaluated in
the present work, we expect it to be in agreement with the one in
Ref.~\cite{Schmidt2012}, similar to what happens with the quasi-particle
residue. 
When compared to the data for short-range interactions from
Ref.~\cite{Schmidt2012} (not shown), the effective mass of the dipolar model
appears to be less affected by interactions and
remains closer to its non-interacting limit ($m^*=m$), in analogy with what
observed for the quasi-particle residue.

It is worth noticing that, through the knowledge of the effective mass, we can
also access the excitation spectrum of the polaron at low momenta, 
\begin{equation}
\epsilon_p(k) = \epsilon_p(k=0) + \frac{\hbar^2}{2 m^*}k^2 + \mathcal{O}(k^4),
\label{eq.ek}
\end{equation}
where $\epsilon_p(k=0)$ is the chemical potential of the polaron discussed in 
Sec.~\ref{subsec.energy}.

\section{Discussion and conclusions}
\label{sect:conclusions}

By means of the DMC method, we have calculated the energetic and structural
properties of a repulsive polaron in a 2D polarized Fermi system. The use of two
different interparticle interactions between the polaron and the bath allows us to analyze the
range of universality of the polaron properties in terms of the gas parameter
$na_s^2$. We show that the polaron energy is universal up to
$na_s^2\approx10^{-5}$; beyond this regime, it depends on the specific shape of
the interaction. Note that our two models also differ for the bath properties
(non-interacting {\it vs.} interacting), which could contribute to this
non-universality.

The presence of a polaron also affects local properties of the bath. The estimation of the 
two-body radial distribution functions and of the static structure factor helps
quantifying this effect. The limit $k\to0$ of the static structure factor gives the excess volume coefficient. Our results for the dipolar model show that the effective volume 
occupied by the impurity shrinks with respect to the one of a particle in the bath. The reason underlying this result is the lack of Fermi correlations between the 
polaron and the medium. On the other hand, for the hard-disk model, in the regime where the exclusion of volume caused by the potential dominates over the Fermi repulsion between bath particles, the volume coefficient becomes positive.

In the weakly-interacting regime, where the effective mass is close to the bare mass of 
the impurity and the quasi-particle residue is the main contribution to the 
ground 
state wave function, the quasi-particle picture is valid. This allows us to 
approximately describe the problem as a quasi-particle made up of the impurity 
``dressed" by the interactions with the bath, propagating through the medium 
with a definite effective mass that takes into account interaction effects.  
When the $Z$ residue starts to depart significantly from one, the 
quasi-particle picture is not able to describe completely the many-body physics 
involved in the problem.

Recent experimental data \cite{Oppong2019} have been reported
for the same system explored here. However, the values of the
gas parameter at which those measurements have been carried out are larger than the universality 
limit determined in this work. Therefore, finite-range effects should be taken 
into account in future theoretical studies to allow for a quantitative comparison with experiments. One can expect that, by fixing both the s-wave scattering length and the effective range, it would be possible to extend the regime of  universality to larger values of the gas parameter. This extension of the regime of universality was recently reported for ultradilute Bose-Bose mixtures \cite{Cikojevic2019}

Data and additional details about the numerical simulations are made publicly
available \cite{Bombin2019Zenodo}.

\acknowledgments{
We acknowledge Richard Schmidt  for providing the data from Ref.~\cite{Schmidt2012}. 
This work has been supported by the Ministerio de
Economia, Industria y Competitividad (MINECO, Spain) under grant No. 
FIS2017-84114-C2-1-P, and by Provincia Autonoma di Trento. Gianluca Bertaina acknowledges D.E. Galli for the access to computational resources at the Department of Physics of the University of Milan.
}

\bibliography{refs}

\end{document}